\documentclass[aps,prl,preprint,groupedaddress,a4paper]{revtex4}

\usepackage{graphicx}

\pdfoutput=1 

\newcommand{\ket}[1]{\left|#1\right>}

\begin{document}


\title{Breakdown of the nuclear spin temperature approach in quantum dot demagnetization experiments}


\author{P. Maletinsky}
\email[]{patrickm@phys.ethz.ch}
\author{M. Kroner}
\author{A. Imamoglu}
\email[]{imamoglu@phys.ethz.ch}
\affiliation{Institute of Quantum Electronics, ETH-Zurich, CH-8093, Zurich, Switzerland}

\date{\today}

\begin{abstract}
\textbf{The physics of interacting nuclear spins arranged in  a crystalline lattice is typically described using a
thermodynamic framework\,\cite{Goldman1970}: a variety of experimental studies in bulk solid-state systems have proven
the concept of a spin temperature to be not only correct\,\cite{Abragam1957,Slichter1961} but also vital for the
understanding of experimental observations\,\cite{Purcell1951}. Using demagnetization experiments we demonstrate that
the mesoscopic nuclear spin ensemble of a quantum dot (QD) can in general not be described by a spin temperature. We
associate the observed deviations from a thermal spin state with the presence of strong quadrupolar interactions
within the QD that cause significant anharmonicity in the spectrum of the nuclear spins. Strain-induced, inhomogeneous
quadrupolar shifts also lead to a complete suppression of angular momentum exchange between the nuclear spin ensemble
and its environment, resulting in nuclear spin relaxation times exceeding an hour. Remarkably, the position dependent
axes of quadrupolar interactions render magnetic field sweeps inherently non-adiabatic, thereby causing an
irreversible loss of nuclear spin polarization.}
\end{abstract}

\maketitle

The study of nuclear spin physics by optical orientation experiments in bulk semiconductor materials has been an
active field of research over the last decades\,\cite{Meier1984,Dyakonov1974,Paget1982}. These research efforts have
shown that using the electron as a mediator, it is possible to transfer angular momentum from light onto nuclei,
thereby establishing a nuclear spin polarization that is orders of magnitude higher than the equilibrium nuclear
polarization at cryogenic temperatures. As a result, the effective nuclear spin temperature in such an optically
pumped system can be pushed to the low mK regime. Combining these optical pumping schemes with nuclear adiabatic
demagnetization techniques borrowed from bulk NMR experiments\,\cite{Slichter1961} would be a natural extension to
these experiments that could lead to a significant further reduction of the nuclear spin temperature. This approach,
previously demonstrated in bulk semiconductors\,\cite{Kalevich1982,Meier1984}, suffers from the fact that in most
systems where optical orientation of nuclear spins is possible, nuclear spin relaxation is too fast to allow for a
significant reduction of magnetic fields in an adiabatic way. Here, we use the exceedingly long nuclear spin
relaxation time in self-assembled QDs\,\cite{Maletinsky2007b} to implement an ``adiabatic'' demagnetization experiment
on the system of $\sim 10^5$ nuclear spins.

The mesoscopic ensemble of nuclear spins in a QD can be conveniently polarized and measured by optical
means\,\cite{Meier1984,Gammon1997,Eble2006,Maletinsky2007a,Maletinsky2007b}. To this end, we use the photoluminescence
(PL) of the negatively charged exciton ($X^{-1}$) under resonant excitation of an excited QD state. It has been shown
previously\,\cite{Lai2006} that under appropriate excitation conditions, $20-50\%$  of the QD nuclear spins can be
efficiently polarized in a timescale of a few ms. The resulting dynamical nuclear spin polarization (DNSP) can then be
measured through a change in the Zeeman splitting, $\Delta E_{\rm OS}$, of the $X^{-1}$ recombination
line\,\cite{Lai2006}; this energy shift due to the spin polarized nuclei is commonly referred to as the Overhauser
shift (OS)\,\cite{Overhauser1953}.

A remarkable feature of the QD nuclear spin  system is the excellent isolation from its environment if the QD is
uncharged. Fig.\,\ref{FigAdiabatDemag}a shows the corresponding free evolution of $P_{\rm nuc}$ (proportional to
$\Delta E_{\rm OS}$) in a QD subject to an external magnetic field $B_{\rm ext}=2~$T. The nuclear spin relaxation time
clearly exceeds one hour and does not vary appreciably over the magnetic field range relevant to this
work\,\cite{Maletinsky2007b}. Since the bulk material surrounding the QD remains unpolarized during the experiment
(see Methods), the long nuclear spin lifetime indicates that nuclear spin diffusion between the QD and its environment
is strongly suppressed. We attribute this quenching of spin diffusion to the structural and chemical mismatch between
the InGaAs QD and its GaAs sourrounding\,\cite{Maletinsky2007a,Malinowski2001}. The very slow nuclear spin relaxation
leaves room for further manipulation of the QD nuclear spin system after optical pumping. In particular, we can study
how $P_{\rm nuc}$ behaves under slow variations of external parameters and thereby study the validity of spin
thermodynamics for the QD nuclear spin system.

If the QD nuclei were describable using a thermodynamic approach, $P_{\rm nuc}$ would be aligned with the external
magnetic field $B_{\rm ext}$ and would be described by Curie's law $\gamma P_{\rm nuc}=B_{\rm ext}C/T_{\rm
spin}$\,\cite{Slichter1961} (here, $\gamma$ is the nuclear gyromagnetic ratio, $C$ the Curie constant and $T_{\rm
spin}$ the nuclear spin temperature). An adiabatic lowering of $B_{\rm ext}$ from an  initial value $B_{\rm i}$ to a
final value $B_{\rm f}$, would conserve $P_{\rm nuc}$ and lead to a reduction of $T_{\rm spin}$ by a factor $B_{\rm
f}/B_{\rm i}$. In general, cooling by adiabatic demagnetization is possible for any system where the spin entropy $S$
is conserved and a function of $B_{\rm ext}/T_{\rm spin}$ only. The ultimate limit to the achievable cooling is
determined by nuclear spin interactions which give the dominant contribution to $S$ at low magnetic fields. The
strength and nature of these interactions can be phenomenologically described by a random local magnetic field $B_{\rm
loc}$. In most cases, $B_{\rm loc}$ is given by the nuclear dipolar couplings ($\approx 0.1~$mT). As soon as $B_{\rm
ext} \approx B_{\rm loc}$, the local fields randomize an established nuclear spin polarization and thereby limit the
efficiency of the adiabatic spin cooling to $B_{\rm loc}/B_{\rm i}$. The resulting behavior of nuclear spin
temperature and polarization as a function of $B_{\rm f}$ is sketched in Fig.\,\ref{FigAdiabatDemag}b: for $B_{\rm
ext}=0$, the spin temperature remains finite and the nuclear spins are completely depolarized. Amazingly, this
depolarization is a reversible process, provided that $S$ is a conserved quantity at all fields. When the spins are
re-magnetized to a magnetic field exceeding $B_{\rm loc}$, their polarization recovers along the direction of the
magnetic field and in particular conserves the sign of its initial spin temperature.

To test the validity of spin thermodynamics  for the QD nuclear spins and to study the possibility of adiabatic
cooling in this system, we performed demagnetization experiments on a QD, as illustrated in
Fig.\,\ref{FigAdiabatDemag}c. A circularly polarized ``pump'' pulse of length $\tau_{\rm pump}$ is used to polarize
the nuclear spins. We then linearly ramp $B_{\rm ext}$ from $B_{\rm i}$ to $B_{\rm f}$ with a rate $\gamma_{\rm
B}=10~$mT/s. At the final field $B_{\rm f}$, the remaining degree of nuclear spin polarization is measured using a
linearly polarized ``probe'' pulse of length $\tau_{\rm probe}$\,\cite{Maletinsky2007b}. This experiment is repeated
at various values of $B_{\rm f}$ to record the process of ``adiabatic'' (de)magnetization.

Figure\,\ref{FigAdiabatDemag}d shows the result of a demagnetization
experiment performed on the nuclear spin system of an individual QD.
The nuclei are polarized with a pump pulse $\tau_{\rm pump}=300~$ms
at $B_{\rm i}=1~$T and measured at $B_{\rm f}$ with a probe pulse
$\tau_{\rm probe}=5~$ms. At a rough glance, this measurement
qualitatively follows the behavior depicted in
Fig.\,\ref{FigAdiabatDemag}b. A closer inspection however, reveals
significant deviations: upon ramping the external field to $B_{\rm
f}=-1~$T we only recover $63 \%$ of the initial $P_{\rm nuc}$. In
addition, by measuring $P_{\rm nuc}(B_{\rm f})$ we determined the
value of the local field to be $B_{\rm loc}=290~$mT: this value is
about three orders of magnitude larger than typical nuclear dipolar
fields. Finally, we observe that even for $B_{\rm f}=0$, the QD has
a remnant nuclear spin polarization $P_{\rm nuc}^{\rm rem}$. To
verify that we do not induce an unwanted increase of spin entropy by
sweeping $B_{\rm ext}$ too fast, we repeated our experiment for
values of $\gamma_{\rm B}$ of $5, 2.5~{\rm and}~0.8 ~$mT/s (crosses
in Fig.\,\ref{FigAdiabatDemag}d). Within the experimentally
accessible range, $\gamma_{\rm B}$ has no influence on our
observations.

The discrepancy between our experimental findings and the  predictions from a thermodynamical treatment of nuclear
spins becomes even more pronounced if we increase $P_{\rm nuc}(B_{\rm i})$ (which can be achieved by increasing
$B_{\rm i}$\,\cite{Maletinsky2007a,Braun2006}). Fig.\,\ref{FigHysteresis}a shows an experiment where we demagnetize
the polarized nuclear spins starting from $B_{\rm i}=2.2~$T to a final field $B_{\rm f}'=-1~$T (black data points). We
then reverse the sweep direction of the magnetic field and ramp $B_{\rm ext}$ back to $B_{\rm f}$ (gray data points).
This experiment shows a considerable hysteresis of the nuclear spin polarization as a function of $B_{\rm f}$. In
particular, $P_{\rm nuc}^{\rm rem}$ changes sign for the two sweep directions of $B_{\rm ext}$. Furthermore, the
magnitude of $P_{\rm nuc}^{\rm rem}$, resp. the width of the observed hysteresis curve depends linearly on the initial
degree of nuclear spin polarization and on $B_{\rm i}$ (Fig.\,\ref{FigHysteresis}b).

To obtain more information about the source of  irreversibility of $P_{\rm nuc}$ during magnetic
field sweeps, we performed a further experiment, where we optically orient the nuclear spins at
$B_{\rm i}=1~$T, ramp the field to a value $B_{\rm f}'<B_{\rm i}$ and then back to $B_{\rm i}=B_{\rm
f}$ where we measure the remaining degree of nuclear spin polarization. The result of this experiment
(Fig.\,\ref{FigHysteresis}c) indicates that the magnetic field sweeps start to induce
irreversibilities in $P_{\rm nuc}$ as soon as $|B_{\rm ext}| \lesssim B_{\rm loc}\approx300~$mT.

Finally we note that the experimental observations  described here do not depend on the sign of the initial nuclear
spin temperature ($T_{\rm spin,i}$). We have repeated the demagnetization experiments for $T_{\rm spin,i}<0$ (i.e.
$\sigma^-$ laser excitation at $B_{\rm i}>0$, not shown here) and observed values of $B_{\rm loc}$ and $P_{\rm
nuc}^{\rm rem}$ consistent with the measurements presented in Fig.\,\ref{FigAdiabatDemag} and
Fig.\,\ref{FigHysteresis}. These measurements are complicated by the fact that for $T_{\rm spin}<0$, nuclear spin
pumping is rather inefficient\,\cite{Maletinsky2007a}, leading to a low degree of DNSP and therefore a smaller signal
to noise ratio than for $T_{\rm spin}>0$.

The three principal features of our experiments, the existence of $P_{\rm nuc}^{\rm rem}$, the hysteretic behavior of
$P_{\rm nuc}$ and the partial irreversibility of our demagnetization experiment, result from a violation of the
nuclear (Zeeman) spin temperature approximation \,\cite{Goldman1970,Slichter1961}. We explain these features by taking
into account the strong inhomogeneous quadrupolar interactions (QI) of the nuclear spins in a QD
\,\cite{Dzhioev2008,Maletinsky2008,Deng2005}. The self-assembled growth of InGaAs QDs is driven by a strong
lattice-mismatch between InGaAs and its surrounding GaAs matrix, which results in a heavily strained QD lattice. As a
consequence, QD nuclei experience large electric field gradients which couple to the nuclear quadrupolar moment. The
resulting quadrupolar Hamiltonian\,\cite{Slichter1996},
\begin{equation}
\label{EqHQI} \hat{H}_{\rm Q}=\frac{h\nu_{\rm
Q}}{2}(\hat{I}^2_{z'}-\frac{1}{3}I(I+1)),
\end{equation}
is characterized by a nuclear quadrupolar frequency $\nu_{\rm Q}$ (proportional to the local strain at the nuclear
site) and a quadrupolar axis $z'$ (with corresponding unit vector $\mathbf{e}_{z'}$ along the main axis of the local
electric field gradient tensor). $\hat{\mathbf{I}}$ is the nuclear spin angular momentum operator with quantum number
$I$ and $\hat{I}_{z'}=\hat{\mathbf{I}}\cdot \mathbf{e}_{z'}$. For typical strain values of
$2~\%$\,\cite{Williamson1999}, we find $\nu_{\rm Q}\approx 2.8~$MHz for As and $1.2~$MHz for In\,\cite{Sundfors1976}.
For comparison of the interaction strength of $\hat{H}_{\rm Q}$ with a pure nuclear Zeeman Hamiltonian $\hat{H}_{\rm
Z}=\gamma \hat{\mathbf{I}}\cdot \mathbf{B}_{\rm ext}$, it is convenient to express the QI strength by an equivalent
magnetic field $B_{\rm Q}=h\nu_{\rm Q}/\gamma$. For As and In, we find $B_{\rm Q}=388~$mT and $125~$mT, respectively;
the corresponding mean value agrees well with our experimental estimate for $B_{\rm loc}$.

The spectrum of a nuclear spin with quadrupolar frequency $\nu_{\rm Q}$ depends strongly on the angle $\theta$ between
$\mathbf{e}_{z'}$ and the external magnetic field (directed along $\mathbf{e}_{z}$). Figure\,\ref{FigQI}b shows the
Eigenenergies of a nuclear spin with $I=3/2$, as a function of $B_{\rm ext}/B_{\rm Q}$. At $B_{\rm ext}=0$, the
spectrum is governed by $\hat{H}_{\rm Q}$, which pairs the nuclear spin states into doublets with angular momentum
projections $\pm m_{z'}$ on $\mathbf{e}_{z'}$. The doublets are split by an energy $|\hbar
\omega_{m_{z'},m_{z'}+1}|=(m_{z'}+1/2)h\nu_{\rm Q}$, respectively. Conversely, in a high magnetic field, the spectrum
is determined by $\hat{H}_{\rm Z}$ with nuclear angular momentum being quantized along the axis $\mathbf{e}_{z}$. Even
at arbitrarily high fields however, the spectrum is significantly perturbed by $\hat{H}_{\rm Q}$ and never becomes
perfectly harmonic.

We modeled our demagnetization experiment using the steady state
solution of a rate equation for the populations $p_{\ket{m}}$ of
spin states $\ket{m}$, which are mutually coupled through dipolar
interactions (Fig.\,\ref{FigQI}b and c). The nuclear spins are
initialized with a Boltzmann distribution at $B_{\rm ext}=B_{\rm i}$
(see Methods) and the evolution of the $p_{\ket{m}}$'s is calculated
as a function of $B_{\rm ext}$. Due to the unequal nuclear spin
level spacings, only nuclear spin flip-flops that preserve
$p_{\ket{m}}$ ($\forall \ket{m}$) are energetically allowed in
general and therefore the spin populations remain invariant as a
function of $B_{\rm ext}$. Varying $B_{\rm ext}$ will change the
relative nuclear spin level spacings in the nonlinear way depicted
in Fig.\,\ref{FigQI}c. Since the $p_{\ket{m}}$'s remain invariant as
$B_{\rm ext}$ is reduced, the nuclear spins are driven into a state
which is out of thermal equilibrium (i.e. not
Boltzmann-distributed). At specific values of $B_{\rm ext}$ (red
markers in Fig.\,\ref{FigQI}c), transition energies between distinct
pairs of nuclear spin states can coincide --- a situation denoted as
a ``cross-over'' of nuclear spin transitions\,\cite{Goldman1970}. At
those fields, the $p_{\ket{m}}$'s are no longer constant and the
nuclear spin levels involved in the cross-over can relax to a
Boltzmann distribution. The irreversibility observed in our magnetic
field sweeps is a consequence of this partial relaxation of nuclear
spins to thermal equilibrium. We speculate that the resulting
increase of the nuclear spin entropy is induced by an
energy-conserving coupling to the environment of the nuclear spins.
If the minimal energy gap of the anti-crossing induced by the
dipolar coupling between two interacting nuclear spins at their
cross-over is smaller than the coupling to the environment, pure
dephasing of the nuclear spin transitions will induce irreversible cross-over transitions and $S$ will
increase.

Upon sweeping $B_{\rm ext}$ through zero (red box in
Fig.\,\ref{FigQI}c), dipolar interactions will couple the states
$m_{z'}=\pm 1/2$. The associated passage through the avoided
crossing between these single-spin states is adiabatic and preserves
the respective populations in the two lowest-lying spin states. In
contrast, nuclear dipolar interaction can not couple any of the
states with $|m_{z'}|>1/2$ due to conservation of energy and angular
momentum. The spin states in the $|m_{z'}| = 3/2$-manifold will
therefore cross and in particular preserve their populations
$p_{3/2}$ and $p_{-3/2}$. The imbalance between these populations
($p_{3/2}<p_{-3/2}$ in Fig.\,\ref{FigQI}) will result in a remnant
polarization $P_{\rm nuc}^{\rm rem}$, even if $B_{\rm ext}$ is
strictly zero.

We averaged our model over a set of parameters $\theta$ and $\nu_{\rm Q}$ to account for the strong inhomogeneity of
QI over the QD (see Methods). The result of this full simulation, is shown in Fig.\,\ref{FigQI}d. We highlight that
the good qualitative agreement with our experimental results (Fig.\,\ref{FigHysteresis}a) is rather insensitive to the
set of parameters used in our simulation. In particular, the choice of the distribution for the parameters $\theta$
and $\nu_{\rm Q}$ did not affect our results significantly. Furthermore, our simulation treats the QD spin system as a
pure spin-$3/2$ system, while for In, $I=9/2$. A numerical treatment of the full InGaAs nuclear spin system is beyond
the scope of this paper and would most likely not alter the qualitative behavior of our simulations (see Methods).

Our results show that the nuclear spin  system of a self-assembled QD provides a rare example for a solid-state
nuclear spin ensemble that can not be described by a nuclear spin temperature\,\cite{Rhim1970}. We note that if one
could assign a spin temperature to the QD nuclear spin system,  optical pumping combined with adiabatic
demagnetization of the nuclear spins would be a novel and efficient way of nuclear spin cooling in QDs without QI:
possible systems include nuclear spin-1/2 systems, such as $^{13}$C-nanotube QDs\,\cite{Marcus2008}, where QI is
inherently absent, or strain-free semiconductor nanostructures \cite{Feng2007}, such as epitaxially grown
droplet-QDs\,\cite{Belhadj2008}. There, adiabatic nuclear spin cooling would only be limited by nuclear dipolar
interactions resulting in $B_{\rm loc}\approx 0.1~$mT. Achieving nuclear spin cooling to temperatures $\approx 100~$nK
should be feasible in these systems, opening ways to studying the remnants of nuclear magnetic phase transitions in
the mesoscopic system of QD nuclear spins \cite{Simon2007}.


\begin{thebibliography}{}

\bibitem{Goldman1970}
Goldman, M., \emph{Spin Temperature and nuclear Magnetic Resonance
in Solids} (Oxford Univ. Press, Oxford, 1970).

\bibitem{Abragam1957}
Abragam, A., and Proctor, W.G., Experiments on Spin Temperature,
\emph{Phys. Rev.}, {\bf 106}, 160-161 (1957).

\bibitem{Slichter1961}
Slichter, C.P., Holton, W.C., and Fellow, A.P.S., Adiabatic
Demagnetization in a Rotating Reference System, \emph{Phys. Rev.},
{\bf 122}, 1701-1708 (1961).

\bibitem{Purcell1951}
Purcell, E.M., and Pound, R.V., A Nuclear Spin System at Negative
Temperature, \emph{Phys. Rev.}, {\bf 81}, 279-280 (1951).

\bibitem{Meier1984}
Meier, F., \emph{Optical orientation}, (North-Holland, Amsterdam,
1984).

\bibitem{Dyakonov1974}
Dyakonov, M.I., and Perel, V.I., Optical Orientation in a System of
Electrons and Lattice Nuclei in Semiconductors - Theory, \emph{Sov.
Phys. JETP}, {\bf 38}, 177-183 (1974).

\bibitem{Paget1982}
Paget, D., Optical-Detection of NMR in High-Purity GaAs - Direct
Study of the Relaxation of Nuclei Close to Shallow Donors,
\emph{Phys. Rev. B}, {\bf 25}, 4444-4451 (1982).


\bibitem{Kalevich1982}
Kalevich, V.K., Kul'kov, V.D., and Fleisher, V.G., Onset of a
nuclear polarization front due to optical spin orientation in a
semiconductor, \emph{JETP Lett.}, {\bf 35}, 20-24 (1982).

\bibitem{Maletinsky2007b}
Maletinsky, P., Badolato, A., and Imamoglu, A., Dynamics of Quantum
Dot Nuclear Spin Polarization Controlled by a Single Electron,
\emph{Phys. Rev. Lett.}, {\bf 99}, 056804 (2007).

\bibitem{Gammon1997}
Gammon, D., Brown, S.W., Snow, E.S., Kennedy, T.A., Katzer, D.S.,
and Park, D., Nuclear Spectroscopy in Single Quantum Dots:
Nanoscopic Raman scattering and Nuclear Magnetic Resonance,
\emph{Science}, {\bf 277}, 85-88 (1997).

\bibitem{Eble2006}
Eble, B., Krebs, O., Lemaitre, A., Kowalik, K., Kudelski, A.,
Voisin, P., Urbaszek, B., Marie, X., and Amand, T., Dynamic nuclear
polarization of a single charge-tunable InAs/GaAs quantum dot,
\emph{Phys. Rev. B}, {\bf 74}, 081306 (2006).

\bibitem{Braun2006}
P.-F. Braun, B. Urbaszek, T. Amand, and X. Marie, Bistability of the nuclear polarization created through optical pumping in In$_{1?x}$Ga$_x$As quantum dots,
\emph{Phys. Rev. B}, {\bf 74}, 245306 (2006).

\bibitem{Maletinsky2007a}
Maletinsky, P., Lai, C.W., Badolato, A., and Imamoglu, A., Nonlinear
dynamics of quantum dot nuclear spins, \emph{Phys. Rev. B}, {\bf
75}, 35409 (2007).

\bibitem{Lai2006}
Lai, C.W., Maletinsky, P., Badolato, A., and Imamoglu, A.,
Knight-field-enabled nuclear spin polarization in single quantum
dots, \emph{Phys. Rev. Lett.}, {\bf 96}, 167403 (2006).

\bibitem{Overhauser1953}
Overhauser, A.W., Polarization of Nuclei in Metals, \emph{Phys.
Rev.}, {\bf 92}, 411-415 (1953).

\bibitem{Malinowski2001}
Malinowski, A., Brand, M.A., and Harley, R.T., Nuclear effects in
unltrafast quantum-well spin-dynamics, \emph{Physica E}, {\bf 10},
13-16 (2001).

\bibitem{Dzhioev2008}
Dzhioev, R.I., and Korenev, V.L., Stabilization of the
Electron-Nuclear Spin Orientation in Quantum Dots by the Nuclear
Quadrupole Interaction, \emph{Phys. Rev. Lett.}, {\bf 99}, 037401
(2007).

\bibitem{Maletinsky2008}
Maletinsky, P., \emph{Polarization and Manipulation of a Mesoscopic
Nuclear Spin Ensemble Using a Single Confined Electron Spin}, (Ph.D.
thesis ETH Z\"urich, Z\"urich, 2008).

\bibitem{Deng2005}
Deng, C.X., and Hu, X.D., Selective dynamic nuclear spin
polarization in a spin-blocked double dot, \emph{Phys. Rev. B}, {\bf
71}, 033307 (2005).

\bibitem{Slichter1996}
Slichter, C.P. \emph{Principles of Magnetic Resonance}, (Springer,
Berlin, 1996).


\bibitem{Vasanelli2002}
A. Vasanelli, R. Ferreira, and G. Bastard,
Continuous Absorption Background and Decoherence in Quantum Dots,
\emph{Phys. Rev. Lett.}, {\bf 89}, 216804 (2002).


\bibitem{Williamson1999}
Williamson, A.J., and Zunger, A., InAs quantum dots: Predicted
electronic structure of free-standing versus GaAs-embedded
structures, \emph{Phys. Rev. B}, {\bf 59}, 15819-15824 (1999).

\bibitem{Sundfors1976}
Sundfors, R.K., Tsui, R.K., and Schwab, C., Experimental gradient
elastic tensors: Measurement in I-VII semiconductors and the ionic
contribution in III-V and I-VII compounds, \emph{Phys. Rev. B}, {\bf
13}, 4504-4508 (1976).

\bibitem{Rhim1970}
Rhim, W.-K., Pines, A., and Waugh, J.S., Violation of the
Spin-Temperature Hypothesis, \emph{Phys. Rev. Lett.}, {\bf 25},
218-220 (1970).

\bibitem{Marcus2008}
Churchill, H.O.H., Bestwick, A.J., Harlow, J.W., Kuemmeth, F.,
Marcos, D., Stwertka, C.H., Watson, S.K., and Marcus, C.M.,
Electron-nuclear interaction in 13C nanotube double quantum dots,
arXiv:0811.3236v2 [cond-mat.mes-hall] (2008).

\bibitem{Feng2007}
Feng, D.H., Akimov, I.A., and Henneberger, F., Nonequilibrium
Nuclear-Electron Spin Dynamics in Semiconductor Quantum Dots,
\emph{Phys. Rev. Lett.}, {\bf 99}, 036604 (2007).

\bibitem{Belhadj2008}
Belhadj, T., Kuroda, T, Simon, C.-M., Amand, T., Mano, T., Sakoda,
K., Koguchi, N., Marie, X., and Urbaszek, B., Optically monitored
nuclear spin dynamics in individual GaAs quantum dots grown by
droplet epitaxy, \emph{Phys. Rev. B}, {\bf 78}, 205325 (2008).

\bibitem{Simon2007}
Simon, P., and  Loss, D.,  Nuclear Spin Ferromagnetic Phase
Transition in an Interacting Two Dimensional Electron Gas,
\emph{Phys. Rev. Lett.}, {\bf 98}, 156401 (2007).

\end{thebibliography}

\section*{Acknowledgments}
We thank A. H\"ogele, J. Elzerman and S.D. Huber for help with the  manuscript, and T. Amand and O. Krebs for fruitful
discussions. We acknowledge A. Badolato for sample-growth. This work is supported by NCCR-Nanoscience.

\section*{Competing Interests}

The authors declare that they have no competing financial interests.

\section*{Correspondence}
Correspondence and requests for materials should be addressed to P.M. and A.I.  (email:
patrickm@phys.ethz.ch, imamoglu@phys.ethz.ch)

\section*{Methods}

\textit{Sample and experimental techniques}

Individual QDs were studied using the photoluminescence (PL) of $X^{-1}$ under resonant excitation of an excited QD
state. The QD sample was grown by molecular beam epitaxy on a $(100)$ semi-insulating GaAs substrate. The approximate
composition of the QDs after self-assembled growth and post-growth annealing was In$_{0.5}$Ga$_{0.5}$As. For
individual optical addressing, the QDs were grown at a low density of $\lesssim 0.1~\mu m^{-2}$. The QDs were spaced
by $25~$nm of GaAs from a doped n$^{++}$-GaAs layer, followed by $30~$nm of GaAs and 29 periods of an AlAs/GaAs
($2/2~$nm) barrier which was capped by $4~$nm of GaAs. A bias voltage applied between the top Schottky and back Ohmic
contacts controls the charging state of the QD. Optical pumping of QD nuclear spins was was performed at the center of
the $X^{-1}$ stability plateau in gate voltage, where PL counts as well as the resulting OS were maximized
\cite{Lai2006}.

\begin{figure}
\includegraphics{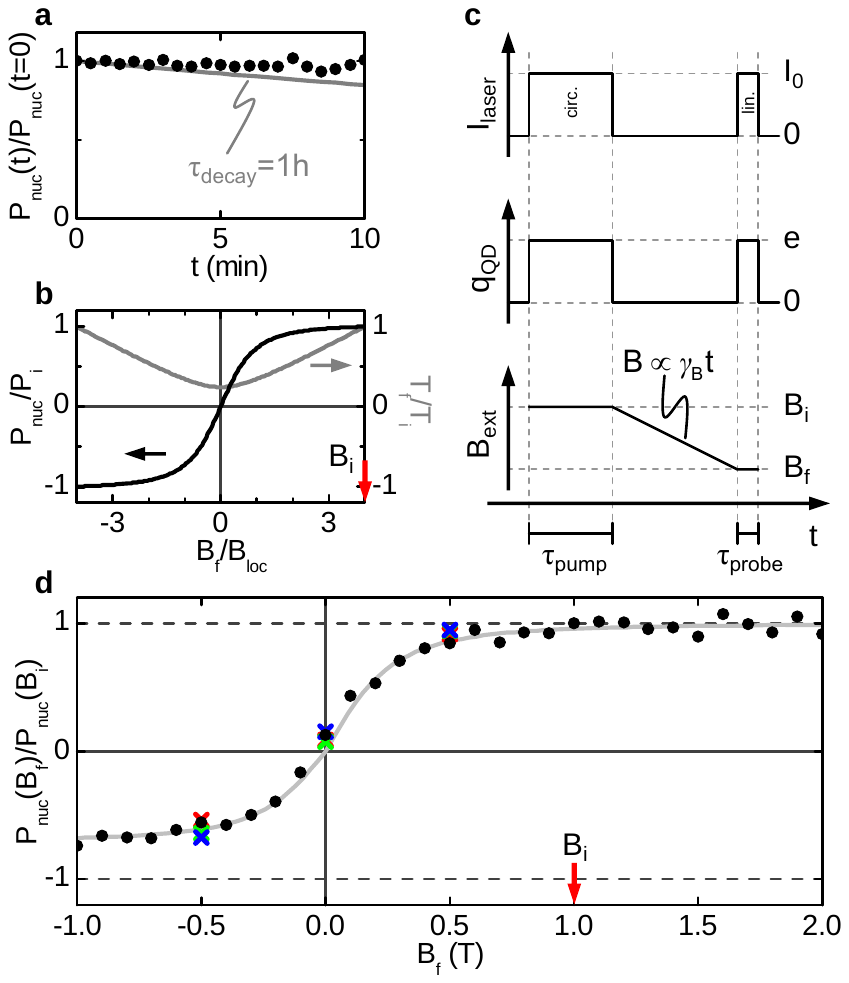}
\caption{\label{FigAdiabatDemag} \textbf{Demagnetization of QD nuclear spins.} (a) Free decay of $P_{\rm nuc}$ at
$B_{\rm ext}=2~$T for an uncharged QD after optical pumping of the nuclear spins for $\tau_{\rm pump}=600~$ms. The
gray curve shows an exponential decay with a time constant of $1~$h for comparison. (b) Theoretical prediction of
nuclear spin temperature and polarization during adiabatic demagnetization from a field $B_{\rm i}$ (red arrow) to
$B_{\rm f}$. (c) Schematic of the experimental procedure for adiabatic demagnetization of QD nuclear spins. The nuclei
are optically pumped at $B_{\rm ext}=B_{\rm i}$ ($T_{\rm spin,i}\sim$ mK). Directly after the pumping pulse, the
electron is ejected from the QD. $B_{\rm ext}$ is then linearly ramped at a rate $\gamma_{\rm B}$ to a value $B_{\rm
f}$ at which we measure $P_{\rm nuc}$. (d) Experimental (de)magnetization of QD nuclear spins. $B_{\rm i}=1~$T as
indicated by the red arrow, $\gamma_B=10~$mT/s and $\Delta E_{\rm OS}(B_{\rm i})=57~\mu$eV. The gray curve is a fit
according to the theoretic predictions shown in (b); we find $B_{\rm loc}=290~$mT. Blue, green and red crosses show a
similar experiment, with $\gamma_B=5$, $2.5$ and $0.8~$mT/s, respectively ($B_{\rm i}=0.5~$T for these data-points). }
\end{figure}

\begin{figure}
\includegraphics[angle=270]{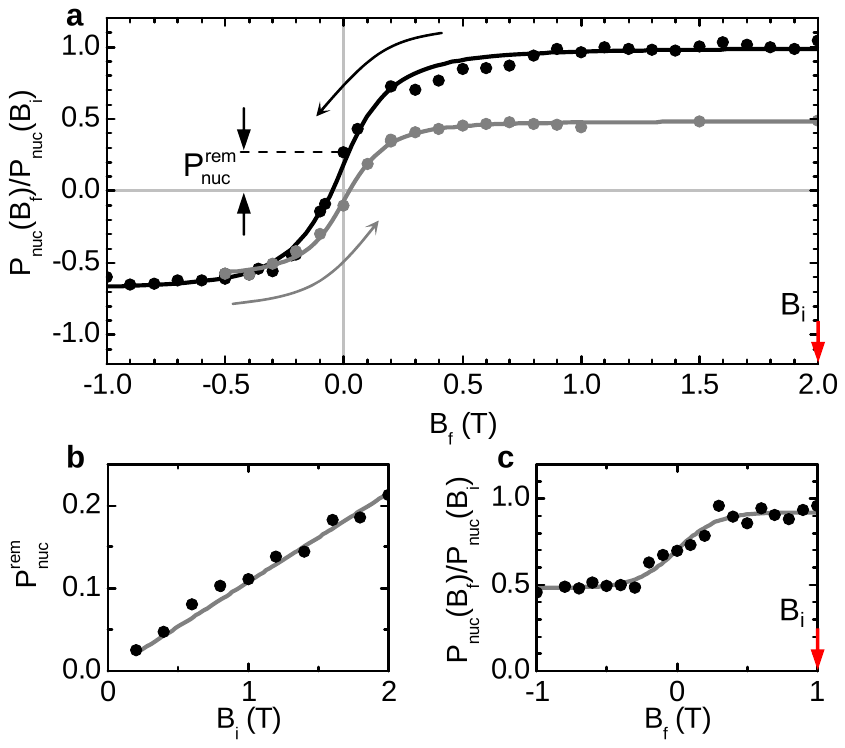}
\caption{\label{FigHysteresis} \textbf{Irreversibility and hysteresis in the demagnetization
experiment}. Black circles: Same experiment as in Fig.\,\ref{FigAdiabatDemag}d, with $B_{\rm
i}=2.2~$T as indicated by the red arrow ($\Delta E_{\rm OS}(B_{\rm i})=89.5~\mu$eV). After reaching
$B_{\rm f}'=-1~$T, we reverse the magnetic field sweep direction and bring the nuclei back to the
initial field (gray circles). (b) The remnant nuclear spin polarization $P^{\rm rem}_{\rm nuc}$
(normalized to the value $P_{\rm nuc}(B_{\rm i})$ found in (a)) as a function of $B_{\rm i}$. As
$P_{\rm nuc}(B_{\rm i})\propto B_{\rm i}$, the nuclear spin temperature after optical pumping is
roughly constant for all values of $B_i$ in this measurement. (c) After polarization of nuclear spins
at $B_{\rm i}=1~$T (red arrow), we sweep $B_{\rm ext}$ to $B_{\rm f}'$ and then back to $B_{\rm i}$,
where  $P_{\rm nuc}$ is measured. The magnetic field sweeps become partly irreversible as soon as
$|B_{\rm f}'| \lesssim 0.3~$T$\approx B_Q$. The lines in the figures are guides to the
eye.}
\end{figure}

The QD sample was immersed in a liquid Helium bath cryostat equipped with a superconducting magnet and was held at the cryostat base temperature of $1.7~$K. The PL emitted by the QD was analyzed in a $750~$mm monochromator allowing for the determination of spectral shifts of the QD emission lines with a precision of $\sim 1~\mu$eV\,\cite{Maletinsky2007a}. A combination of an optical
``pump-probe'' technique, together with linear ramps of the applied magnetic field were used to adiabatically
demagnetize the QD nuclear spins (see Fig.\,\ref{FigAdiabatDemag}c); technical details of the pump-probe
setup are given elsewhere\,\cite{Maletinsky2007b}. The ``pump'' pulse consists of a circularly
polarized laser pulse of duration $\tau_{\rm pump}$ which is used to optically orient the QD nuclear
spins\,\cite{Maletinsky2007a}. We typically achieve an OS of $\Delta E_{\rm OS}=60~\mu$eV at $B_{\rm i}=1~$T, corresponding to nuclear spin polarization $P_{\rm nuc}\approx 35\%$ or $T_{\rm i}\approx 1.5~$mK (for $B_{\rm i}=2.2~$T$, \Delta E_{\rm OS}=89.5~\mu$eV and $P_{\rm nuc}\approx 50\%$). In the range of $B_{\rm ext}$ relevant to our experiment, $P_{\rm nuc} \propto B_{\rm i}$\,\cite{Maletinsky2007a} such that the initial nuclear spin temperature $T_{\rm i}$ is roughly constant and on the order of few mK\,\cite{Dyakonov1974} for all values of $B_{\rm i}$.

Directly after applying the pump pulse to the QD, the gate voltage is switched to a value where the QD is
charge-neutral. In this regime, nuclear spin polarization has an exceedingly long relaxation time on the order of
hours\,\cite{Maletinsky2007b} (see Fig.\,\ref{FigAdiabatDemag}a). We note that we can exclude any significant nuclear
polarization of the bulk material surrounding the QD. The observation of DNSP in our experiment depends sensitively on
the excitation laser energy, which we tune to an intra-dot (p-shell) excitation resonance with a width of $\approx
300~\mu$eV and located $\approx 36~$meV above the PL emission energy. The sharpness and energy of this excitation
resonance makes any excitation processes which involve the creation of free electrons in the bulk very
unlikely\,\cite{Vasanelli2002}. Furthermore, the pumping time $\tau_{\rm pump}=600~$ms used in our experiment is much
too short to lead to a significant bulk nuclear spin polarization, even if some free electrons were created during
laser illumination.

\textit{Details of the model}

The model we developed to explain our experimental findings is based on the steady state solution of a rate equation
for the populations of a nuclear spin $I=3/2$ system.  The nuclear spins are initialized with a Boltzmann distribution
over the spin states at $B_{\rm ext}=B_{\rm i}$. The assumption of a thermal distribution of nuclear spin levels at
$B_{\rm ext}=B_{\rm i}$ is justified by the fact that nuclear spins are polarized by hyperfine interaction with the QD
electron: optical pumping of the electron leads to a broadening  of its spin states by several
$\mu$eV\,\cite{Maletinsky2007a}, allowing for electron-nuclear flip-flops between the electron and any two given
nuclear spin states which are coupled by the hyperfine interaction. It is therefore reasonable to assume that the
occupations of nuclear spin levels at $B_{\rm i}$ follow a Boltzmann distribution.

\begin{figure}
\includegraphics{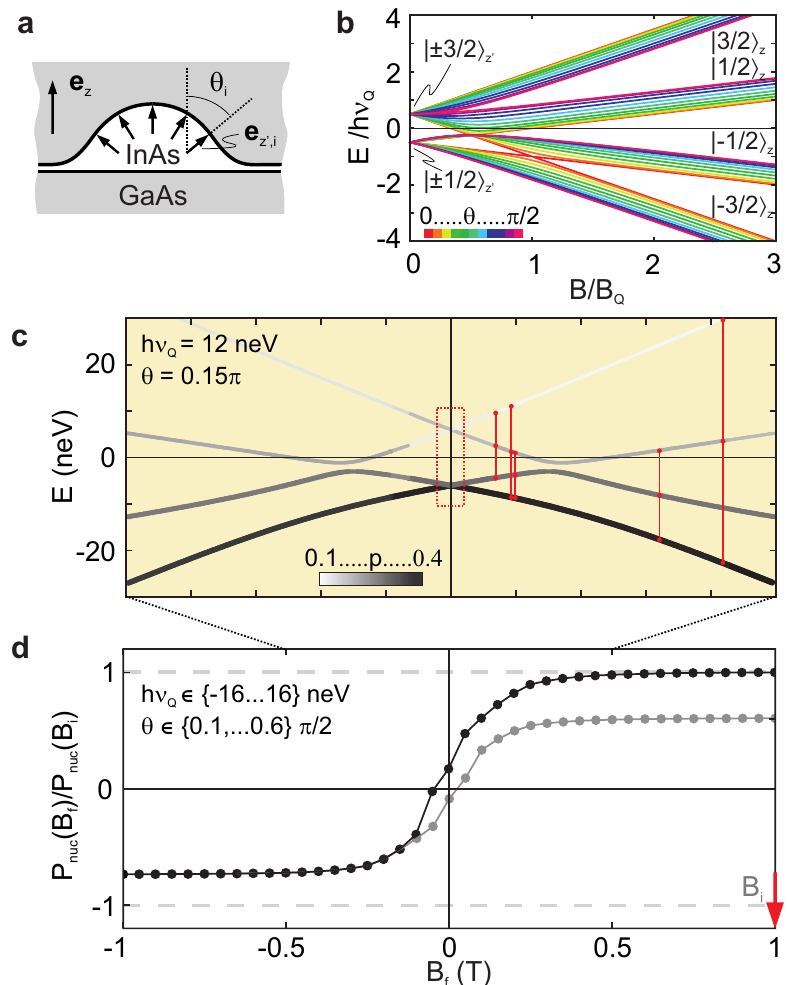} \caption{\label{FigQI} \textbf{Modeling of the demagnetization experiment}. Local
electric field gradients induced by strain in self-assembled QDs result in strong QI for the nuclear spins. (a) Model
of local strain axis-distribution $\mathbf{e}_{z'}$ within a QD. (b) Spectrum of nuclear spins (I=3/2) under the
influence of both $\hat{H}_{\rm Q}$ and $\hat{H}_Z$ for a variety of angles
$\theta$ between $\mathbf{e}_{z'}$ and $\mathbf{e}_{z}$. (c) Simulation of QD nuclear spin
demagnetization for a particular
setting $\nu_{\rm Q}=3~$MHz and $\theta=0.15\pi$. Nuclear spin populations $p$ 
are represented both by line-thickness and grayscale of the lines that indicate the energy of the
nuclear spin states. At $B_{\rm i}=1~$T, the nuclei are initialized with a Boltzmann distribution
over their spin states. The populations remain constant for most values of $B_{\rm f}$. Only if a
cross-over of nuclear spin transitions occurs (red markers for $B_{\rm f}>0$), the occupations of the
involved spin states evolve to a (local) thermal equilibrium distribution (see text). We simulate
this process for a set of configurations $\{\theta,\nu_{\rm Q}\}$ and calculate the corresponding
magnetization $P_{\rm nuc}\propto \langle I_z \rangle$. (d) shows the resulting nuclear spin
polarization as a function of $B_{\rm f}$ starting at $B_{\rm i}$ (red arrow), which qualitatively
reproduces the experimental findings shown in Fig.\,\ref{FigHysteresis}.}
\end{figure}

We then change the magnetic field by keeping the populations of  spin levels fixed. Only at the
specific fields where cross-relaxation is permitted (Fig.\,\ref{FigQI}c), we allow for a
local thermal equilibrium to be established between the spin levels involved in the cross-relaxation
transitions. All other populations and the total energy of the nuclear spin system remain constant.
Upon sweeping through $B_{\rm ext}=0$, we assume that the levels $m_{z'}=\pm1/2$ undergo an adiabatic
passage through an anticrossing induced by the coupling of these two states by dipolar interactions.
Spin states with $m_{z'}=\pm3/2$ however remain uncoupled and undergo an adiabatic level crossing
which preserves their populations.

The result of our simulations is shown in Fig.\,\ref{FigQI}c and d of the main text. We illustrate the
evolution of the occupations of the individual nuclear spin states in Fig.\,\ref{FigQI}c, where we show the
spectrum of a nuclear spin for the parameters $\nu_{\rm Q}=3~$MHz, $\theta=0.3\pi/2$ and
$\gamma=10~$MHz/T. The occupations of the individual levels is encoded by the thickness and gray
shade of the corresponding lines. Magnetic fields where cross-relaxation processes take place are
indicated by red lines. We repeated this calculation for a set of angles $\theta \in\{ 0.1,0.2,...,0.6\}
\frac{\pi}{2}$ and quadrupolar frequencies $\nu_{\rm Q} \in \{ -4,-3,-2,2,3,4\}~$MHz\, over which we
average our results. Since the local strain in our QDs can be both tensile and compressive, positive
and negative values for $\nu_{\rm Q}$ are possible. By solving the complete Hamiltonian
$\hat{H}_Q+\hat{H}_{\rm Z}$, we can relate the occupancies of the spin levels to our experimentally
observed nuclear spin polarization --- the expectation value $\langle \hat{I}_z \rangle$ of the
nuclear spin polarization along the direction of $B_{\rm ext}$. Fig.\,\ref{FigQI}d of the main paper shows the
result of our simulation in form of the calculated evolution of $P_{\rm nuc}$ as a function of
$B_{\rm f}$.

We note that our model is a great simplification of the actual  experimental situation. First, we completely neglect
cross-relaxation events between nuclei of different $(\theta,\nu_{\rm Q})-$values. Second, our calculation was
performed for a spin-$3/2$ system for simplicity, while the actual QD nuclear spin system consists of a mixture of
spin-$3/2$ (Ga, As) and spin-$9/2$ (In), which further complicates the situation. While a numerical treatment of the
full InGaAs nuclear spin system is beyond the scope of this paper, we argue that such a treatment would not alter the
physical picture conveyed by our simulation. Including I=9/2 spins would lead to a nuclear spin spectrum similar to
the one illustrated in Fig.\,\ref{FigQI}b. The number of magnetic field values where cross-relaxation events would be
energetically allowed would increase compared to the case of I=3/2, but these events would still be singular in the
sense that for most values of $B_{\rm ext}$, the nuclear spins could not thermalize. The system would thus still be
driven out of thermal equilibrium and the relaxation events during cross-relaxation would lead to an increase of
nuclear spin entropy. Including flip-flop events between In and As nuclear spins would have a similar effect: these
transitions would be allowed for a subset of close nuclei and would allow for partial thermalization only at specific
values of $B_{\rm ext}$.

\end{document}